\documentclass [aps,prb,10pt,twocolumn,showpacs,superscriptaddress]{revtex4-1}
\usepackage[english]{babel}
\usepackage{graphicx}
\usepackage{amsmath,amssymb}
\usepackage[utf8]{inputenc}
\usepackage{natbib}
\usepackage{units}
\usepackage{hyperref}
\usepackage{color}
\usepackage{url}
\hypersetup{colorlinks=true,citecolor={blue},linkcolor={blue},urlcolor={blue}}
\usepackage{balance}
\usepackage{textcomp}
\begin{document}
\title{Parity solitons in nonresonantly driven-dissipative condensate channels}

\date{\today}

\author{H. Sigurdsson}
\email[correspondence address:~]{helg@hi.is}
\affiliation{Science Institute, University of Iceland, Dunhagi-3, IS-107 Reykjavik, Iceland}
\author{T.C.H. Liew}
\affiliation{Division of Physics and Applied Physics, School of Physical and Mathematical Sciences, Nanyang Technological University 637371, Singapore}
\author{I.A. Shelykh}
\affiliation{Science Institute, University of Iceland, Dunhagi-3, IS-107 Reykjavik, Iceland}
\affiliation{ITMO University, St.\ Petersburg 197101, Russia}

\begin{abstract}
We study analytically and numerically the condensation of a driven-dissipative exciton-polariton system using symmetric nonresonant pumping geometries. We show that the lowest condensation threshold solution carries a definite parity as a consequence of the symmetric excitation profile. At higher pump intensities competition between the two parities can result in critical quenching of one and saturation of the other. Using long pump channels, we show that the competition of the condensate parities gives rise to a new type of topologically stable defect propagating indefinitely along the condensate. The defects display repulsive interactions and are characterized by a sustained wavepacket carrying a pair of opposite parity domain walls in the condensate channel.
\end{abstract}

\maketitle
\section{Introduction}
A great deal of work has been devoted in understanding the physics of equilibrium condensate systems such as cold atoms and superconductors. Within the mean field theory the nonlinear nature of these quantum fluids is eloquently captured revealing superfluid currents, vortices, and solitons~\cite{pitaevskii_2003bose}. Solitons are self-supporting wavepackets maintaining their shape and group velocity as a consequence of dispersive and nonlinear terms compensating each other. They have been studied and observed in numerous physical systems such as optical media~\cite{Kivshar_OpticalSolitons_Sci2003}, proteins~\cite{davydov1985solitons}, superfluids~\cite{Maki1977}, Bose-Einstein condensates~\cite{Becker_AtOscDarkBrightSol_NatPhys2008}, and magnetic materials~\cite{Kosevich1998}. They can be classified as non-topological or topological, the latter meaning that they belong to a group of different homotopy than the soliton-free state and thus are stable against decay to a topologically trivial field distribution. 

One should also distinguish between conservative solitons appearing in the systems described by the nonlinear Schrödinger, Korteweg- de Vries and sine-Gordon equations and dissipative solitons appearing in the systems described by various modifications of the complex Ginzburg-Landau equation~\cite{Lorenz_GinzburgLandauEQ_RMP2002, keeling_spontaneous_2008} (also referred as the generalized Gross-Pitaevskii equation in the context of BEC). The complex Ginzburg-Landau equation is a powerful tool to understand wave phenomena in diffusive nonlinear systems and has successfully predicted the existance of various defects, chaos, turbulence, bifurcation, with solutions from traveling waves to Nozaki-Bekki holes. Dissipative solitons are amongst these solutions and can exist in exciton-polariton condensates~\cite{Littlewood_DissCond_PRL2006, Littlewood_DissCond_PRB2007, carusotto_quantum_2013}, optical parametric oscillators~\cite{Wouters_GoldStoneOPO_PRA2007, Yamamoto_OPOnetwork_Nature2014}, cold atoms~\cite{Helmut_coldatoms_RevMod2013},  and optically driven Rydberg clusters~\cite{Saffman_RydbergOpt_RevMod2010}. In all these cases a dissipative macroscopic quantum state is continuously replenished by external driving. For this, exciton-polaritons are excellent candidates displaying the solid state analog of Bose-Einstein condensation under either optical or electrical driving~\cite{kasprzak_bose-einstein_2006, balili_bose-einstein_2007, lai_coherent_2007, carusotto_quantum_2013, Schneider_ElecPolLas_Nature2013, Byrnes2014} for surprisingly high temperatures~\cite{Guillaume_RoomTempPol_APL2002, Christop_RoomTempCond_PRL2007}. This opens a way for the potential application of polaritonic systems in design of optoelectronic devices of next generation~\cite{Fraser2016, Sanvitto2016}. Indeed, polariton solitons~\cite{Skryabin_PolSolitons_CRP2016} have been considered as candidates for information processing schemes~\cite{Cancellieri_SolitonGate_PRB2015}, are compatible with topological polariton systems~\cite{Kartashov_TopSoliton_Optica2016} and their entanglement has been suggested~\cite{Zhang_VecSoliton_OpEx2010}.

Due to strong polariton-polariton interactions polariton condensates also represent a unique laboratory for the simulation of a plethora of nonlinear phenomena. Features of dark and bright solitons, although not shown to stay supported indefinitely, were recently observed in phase locked polariton condensates~\cite{Berloff_OscSolitons_NJP2014} and have been predicted in hyperbolic regions of negative effective mass in patterned microcavities~\cite{Arnardottir_NanoWHyp_ACS2016}. Dark solitons~\cite{Grosso_VortexStreets_PRL2011} were found to eventually relax due to the dissipative physics of nonequilibrium condensates~\cite{ Smirnov_DarkSolitons_PRB2014} with the exception of trapping~\cite{Elena_PerPotSol_PRL2013} but with no evidence as of yet for propagation. The prediction of oblique dark solitons~\cite{ElG_ODS_PRL2006} was also verified for polariton fluids~\cite{Amo_PolOblSol_Sci2011} followed by the prediction of oblique dark half-solitons in spinor condensates~\cite{Flayac_ODHS_PRB2011} and later their controversial experimental observation~\cite{Hivet_HalfSol_NatPhys2012, Cilibrizzi_DarkSolitons_PRL2014, Amo_DarkSolComment_PRL2015, Cilibrizzi_DarkSolreply_PRL2015}. Furthermore, dissipative solitons~\cite{Elena_DissSoliton_PRA2012} and bright solitons have been predicted~\cite{Egorov_BrigthSoliton_PRL2009} with the latter observed~\cite{SichM_BrightSoliton_Nat2012} in polariton fluids. 

In this paper, we analyze the gain and dissipation properties of a polariton condensate under nonresonant CW symmetric spatial pumping. Many properties of the complex Ginzburg-Landau equation are well studied under uniform driving \cite{Lorenz_GinzburgLandauEQ_RMP2002} but in experiment the use of symmetric pump shapes is a conventional protocol and thus deserves some investigation. We show that in 1D systems the condensation threshold is determined by an order parameter of definite parity due to the symmetric coordinate dependent nature of the pump gain. To the best of our knowledge, only special instances on how parity relates to the condensation threshold have been studied~\cite{Ohadi_PRX2016}. We also show that a second critical pump intensity exists where the uncondensed parity suddenly condenses and drives the existing parity to zero, an effect best described as parity cross-saturation. Extending the 1D system to 2D opens up a new spatial degree of freedom allowing polaritons to travel parallel to the pump where we observe a type of topologically distinct defect state traveling without dissipation in the condensate. The defects exist only in the presence of nonlinearities, give rise to nontrivial currents, and possess a pair of domain walls of opposite parity to the defect free condensate, making them {\it parity solitons}. They are found to exist over a wide range of pump powers, pump shapes, nonlinearities and even with no pump induced- or external trapping. Our work shows propagating non-dissipative soliton states in nonresonantly driven polariton condensates -- an important step towards realizing optoelectronic platforms based on soliton signals.

\section{Theory}
Spinless driven-dissipative polariton condensates can be accurately modeled using the complex Ginzburg-Landau equation for the scalar order parameter $\Psi$~\cite{wouters_excitations_2007, keeling_spontaneous_2008, wouters_energy_2010}. 
\begin{equation} \label{eq.GP0}
i \dot{\Psi} = \left[ - \frac{\hbar \nabla^2}{2m} +  (g_P + i) P f(\mathbf{r}) - i\Gamma +  (\alpha - i R) |\Psi|^2\right] \Psi.
\end{equation}
Here $f(\mathbf{r})$ and $P$ are the nonresonant pump profile and intensity, $g_P$ is the exciton reservoir blueshift induced by the pump, $m$ is the polariton mass, $\Gamma$ is the polariton decay rate, $\alpha>0$ accounts for polariton-polariton interactions (de-focusing), and $R$ is the saturation rate. 

\begin{figure}[!t]
  \centering
		\includegraphics[width=0.5\textwidth]{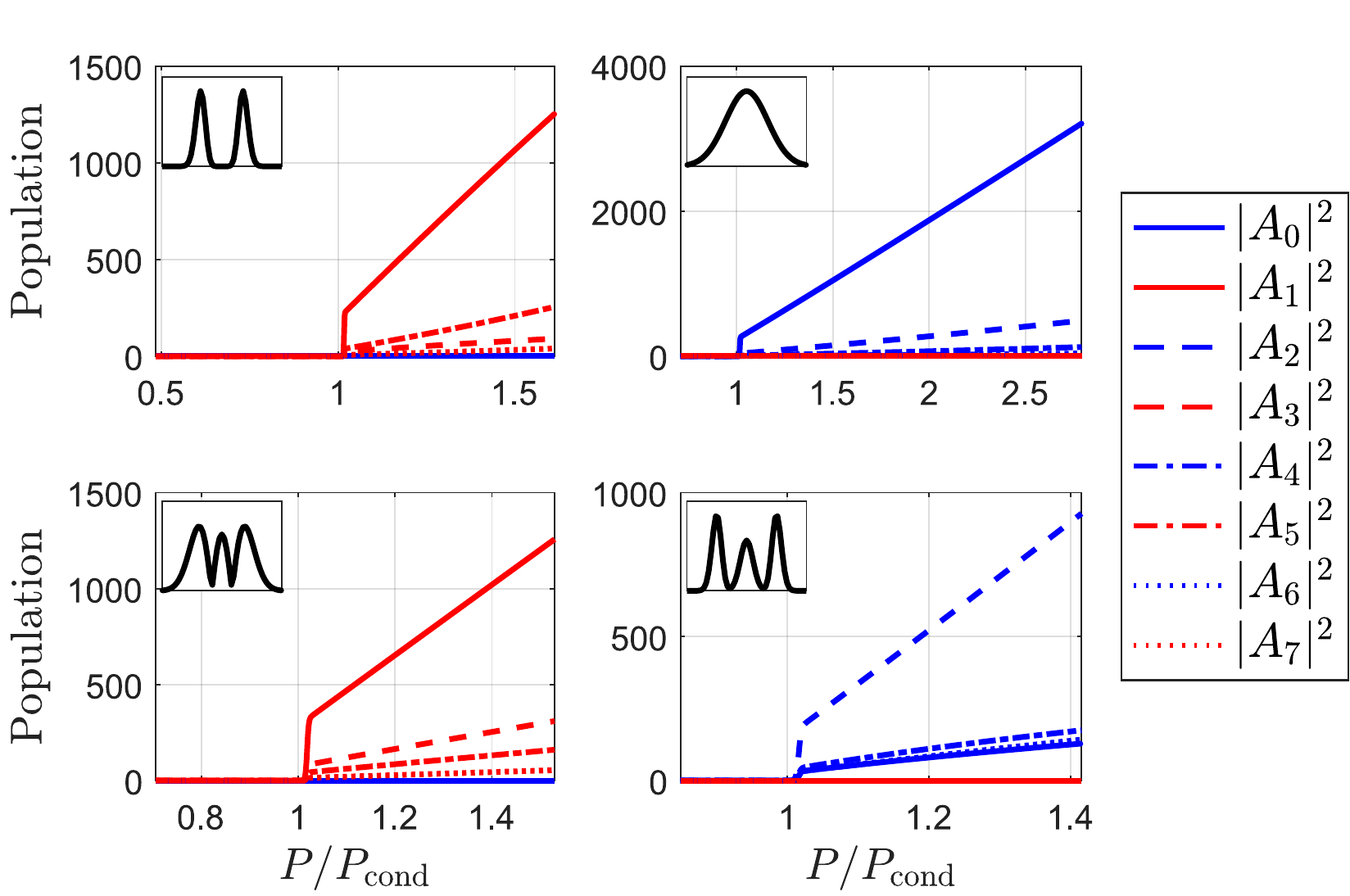}
	\caption{Four separate threshold results for a 1D condensate driven by a symmetric pump profile $f(x)$ (inset) resulting in a definite parity condensate. Here $\alpha = g_P = 0$.}
\label{fig1}
\end{figure}

We begin our analysis on a symmetrically excited condensate in 1D geometry ($f(x) = f(-x)$) and setting $\alpha = g_P = 0$. The dynamics of the order parameter is then characterized only by the dispersion and gain-dissipation mechanics. By slowly ramping the pump intensity polaritons condense at $P = P_\text{cond}$ into a lowest threshold solution with definite parity. Since the condensate decays quickly to zero as polaritons move away from the pump we can choose the infinite quantum well basis $\{\psi_n\}$ with boundaries $|x| = L/2$ far away from the condensate in order to extract the parity dependent behavior of the condensate, the order parameter is then written $\Psi = \sum_n A_n(t) \psi_n(x)$. Integrating out the coordinate dependence we get:
\begin{align} \notag
 \dot{A_n} =& -\left( i \omega_n +  \Gamma \right) A_n +   P \sum_m f_{nm} A_m  \\ \label{eq.An1}
 &-   R   \sum_{jkl} M_{njkl} A_j^* A_k A_l.
\end{align}
Here $\hbar\omega_n$ are the linear real eigenenergies. The pump elements $f_{nm}$ are written as $f_{nm} = \int f(x) \psi_n \psi_m \ dx$, being nonzero only when the product $\psi_n \psi_m$ is even. The nonlinear elements are written as $M_{ijkl} = \int \psi_i \psi_j \psi_k \psi_l \ dx.$ Defining $n_n \equiv |A_n|^2$ and looking at the rate of the modes in the linear regime $(|\Psi|^2 \sim 0)$ we have,
\begin{equation} \label{eq.rateAn}
 \frac{\dot{n_n}}{2} = P\sum_{m} f_{nm} \sqrt{n_n n_m} \cos{(\phi_n - \phi_m)}  - \Gamma n_n,
\end{equation}
where $\phi_n$ is the phase of $A_n$. This means that the maximum gain of the system is determined by a superposition of same parity modes coupled through $f_{nm}$. This is in agreement with numerical results where $P$ in Eq.~(\ref{eq.GP0}) is adiabatically ramped from weak stochastic initial condition for different pump profiles (see Fig.~\ref{fig1}).

\begin{figure}[!t]
  \centering
		\includegraphics[width=0.5\textwidth]{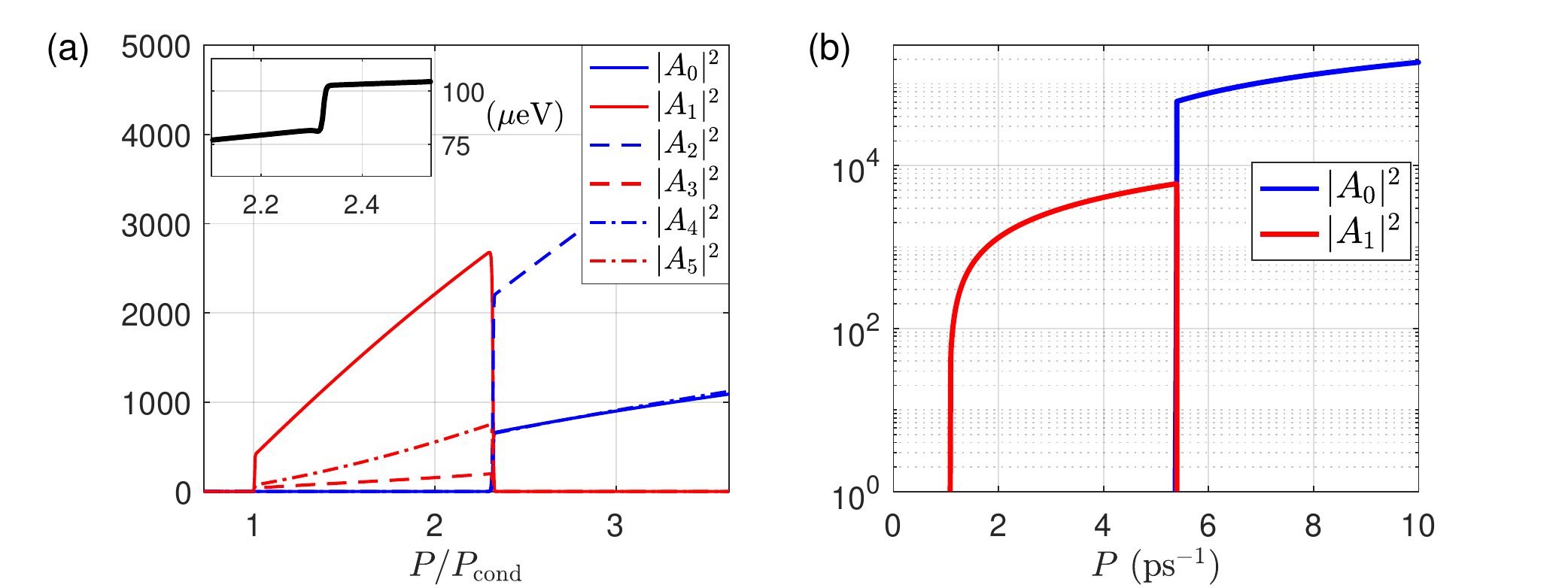}
	\caption{(a) Mode population and condensate energy (inset) as a function of pump power for 1D system showing the parity switch. (b) Eqs.~(\ref{eq.A0}) and (\ref{eq.A1}) propagated with increasing pump power with $\omega_0 = \omega_1 = 0$. Element values were set to $M_{0000} = 0.001$ $\mu$m$^{-1}$, $M_{1111} = 0.0573$ $\mu$m$^{-1}$, $M_{0011} = 0.01$ $\mu$m$^{-1}$, $f_{00} = 0.032$, $f_{11} = 0.095$, $R = 0.0012$ ps$^{-1}$$\mu$m, and $\Gamma = 0.1$ ps$^{-1}$. Eqs.~(\ref{eq.Pcond}) and (\ref{eq.Pcrit}) give $P_\text{cond} = 1.05$ ps$^{-1}$ and $P_\text{crit} = 5.3$ ps$^{-1}$ in good agreement with numerical results.}
\label{fig2}
\end{figure}

As the pump intensity is increased, we observe a second critical pump power $P=P_\text{crit}$ where the condensate suddenly flips its parity followed by a shift in energy (see Fig.~\ref{fig2}a). The effect demonstrates the presence of a second condensate solution of the opposite parity, for which the occupation rate changes from net negative to positive and which in return quenches the first solution. This scenario becomes clear if one uses a truncated basis of linear eigenstates, $\Psi = A_0 \psi_0 + A_1 \psi_1$, which results in the following coupled equations:
\begin{align} \notag 
\frac{\dot{A_0}}{A_0}  =&- (i\omega_0 +  \Gamma) +  P f_{00}  -R \big[M_{0000} |A_0|^2  \\ \label{eq.A0}
& + 2 M_{0011} |A_1|^2 + M_{0011} A_1^2 e^{- i 2\phi_0} \big], \\ \notag
\frac{\dot{A_1}}{A_1}  = &-(i\omega_1 + \Gamma)  +  P f_{11} - R \big[M_{1111} |A_1|^2 \\  \label{eq.A1}
& + 2 M_{0011} |A_0|^2  + M_{0011} A_0^2 e^{- i 2\phi_1} \big].
\end{align}
The rate equations then become,
\begin{align} \label{eq.A0rate}
\frac{1}{2} \frac{\dot{n_0}}{n_0}  =  P f_{00} - \Gamma  -  R \big[ M_{0000} n_0 +  M_{0011} n_1 (2 -  \cos{(2\phi)} ) \big] , \\  \label{eq.A1rate}
\frac{1}{2} \frac{\dot{n_1}}{n_1}  =  P f_{11}  -\Gamma -  R  \big[ M_{1111} n_1  +  M_{0011} n_0 (2 +   \cos{(2\phi)}) \big],
\end{align}
where $\phi = \phi_1 - \phi_0$. The condensation threshold is then determined by,
\begin{equation} \label{eq.Pcond}
P_{\text{cond}} = \Gamma \cdot \min{ \left\{ f_{00}^{-1}, \ f_{11}^{-1} \right\} }.
\end{equation}
Let us assume that $P_\text{cond} = \Gamma/f_{11}$ is minimal. Then the $A_1$ mode condenses first and has a steady state according to $n_1 = (P f_{11} - \Gamma)/R M_{1111}$. Eqs.~(\ref{eq.A0rate}) and (\ref{eq.A1rate}) show that cross-saturation effects are tunable through the phase difference between the two modes. Without any loss of generality we can set $\omega_0 = \omega_1 = 0$ which restricts the phase to $\phi = k \pi/2$ where $k \in \mathbb{Z}$. It becomes then obvious that $k = 0$ creates the optimum condition for the $A_0$ mode to become populated since it minimizes the saturation caused by the $A_1$ mode. The critical pump power where the rate of the $A_0$ mode turns positive is then,
\begin{equation} \label{eq.Pcrit}
P_\text{crit} = \Gamma \frac{1 - M_{0011}/M_{1111}}{f_{00} - M_{0011} f_{11}/M_{1111}}.
\end{equation}
The above expression predicts the condensation of the uncondensed parity (here $A_0$) but does not necessarily guarantee that the existing parity ($A_1$) is driven to zero. The complete quenching of the existing parity corresponds then to a class of solutions determined by the elements $f_{nm}$ and $M_{ijkl}$ where $P \geq P_\text{crit}$ causes the rate of the previously dominant parity to become strictly negative, driving it to zero. If $M_{1111} > M_{0011}$ and $f_{11} \geq f_{00} > f_{00}M_{0011}/M_{1111}$ then one has $P_\text{crit} \geq P_\text{cond}$. For $f_{11}<f_{00}$ the lowest threshold belongs to the other parity, for $f_{00} < f_{00}M_{0011}/M_{1111}$ one has $P_\text{crit}<0$ which has no physical interest. In Fig.~\ref{fig2}b we show the parity flip for Eqs.~(\ref{eq.A0}) and (\ref{eq.A1}) for an arbitrary set of pump and nonlinear elements in good agreement with Eqs.~(\ref{eq.Pcond}) and (\ref{eq.Pcrit}). We explicitly synchronized the energies in order to emphasize that the effect can purely be explained via the gain-decay mechanism. We also note that the effect does not vanish for nonzero real interactions $\alpha$ and/or additional confining potentials $V(x)$, and can be retrieved using the reservoir approach (see Sec.~\ref{sec.app1}). Such pump induced pattern reconfiguration was previously observed in Ref.~[\onlinecite{Ge_PatternForm2013}] but lacking explanation on the underlying mechanism.

In the case of an asymmetric pump profile the pump elements $f_{nm}$ start mixing together the gain of the two parities. We investigate the effect such asymmetry and find that at critical skewing strength the parity switch is replaced by a solution of mixed modes resulting in an asymmetric condensate. The presence of noise in the pump is also investigated and is found to have a small effect on the parity switch (see Sec.~\ref{sec.app2}).
\begin{figure}[!t]
  \centering
		\includegraphics[width=0.5\textwidth]{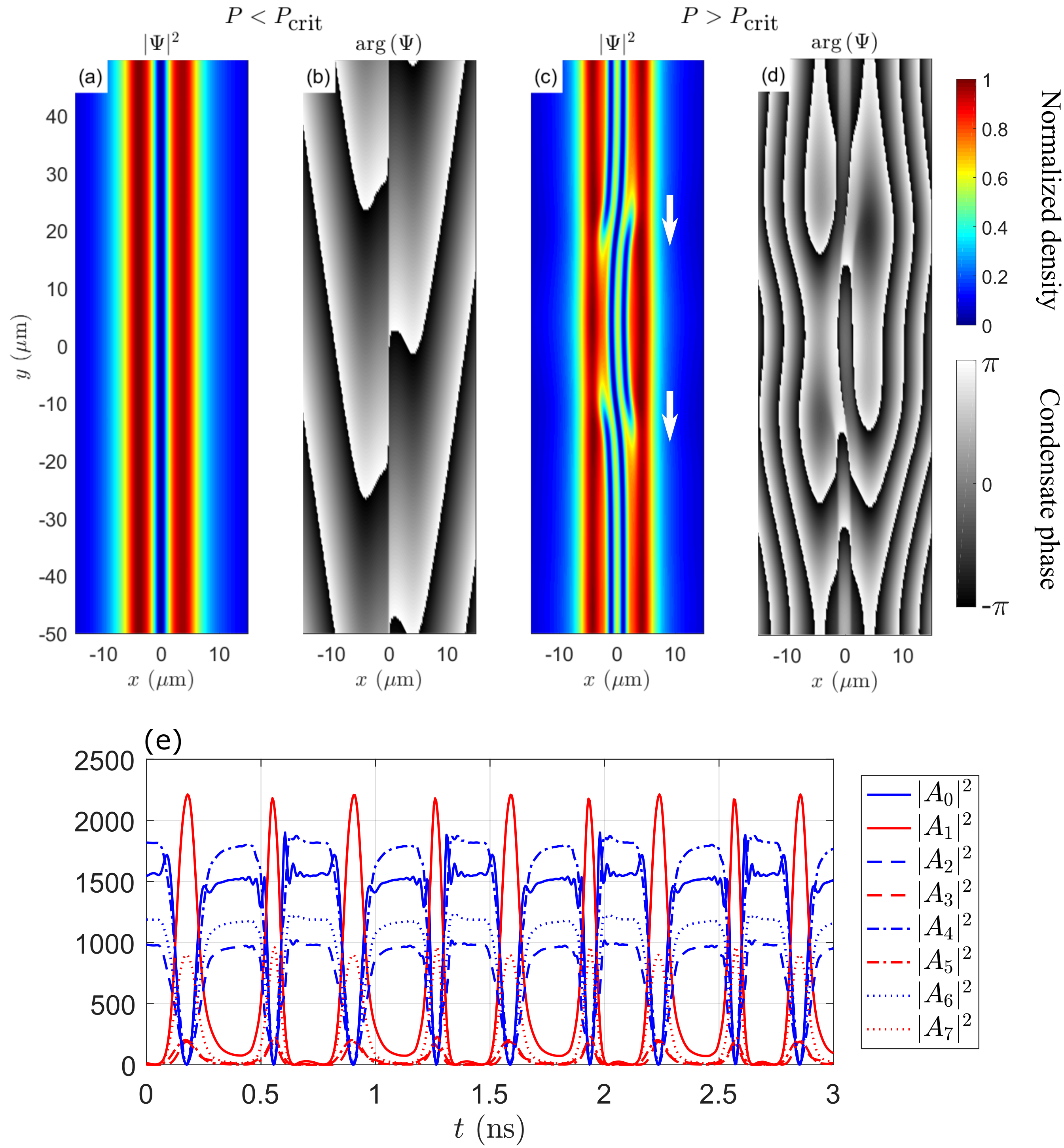}
	\caption{Normalized density (a) and phase (b) of the condensate in an odd parity solution with $k_y \neq 0$ with $P = 1$ ps$^{-1}$. (c,d) The same channel after driving the pump intensity past $P_\text{crit}$. Background noise and finite $k_y$ wavevectors result in the formation of defects traveling along the channel indefinitely (white arrows). Here $P = 3.5$ ps$^{-1}$. (e)  Mode population at $y=\mbox{const.}$ from panels (c,d). The even parity (blue lines) of the channel drops to zero when the odd parity defects travels past. Here the two defects travel at a fixed velocity and without decaying.}
\label{fig3}
\end{figure}

\section{Parity Solitons}
We now return to Eq.~(\ref{eq.GP0}) with $\alpha \neq 0$ and $g_P \neq 0$. Considering the realistic case of 2D exciton-polaritons in planar microcavity structures we chose the pump profile from Fig.~\ref{fig1}a in the form of a channel along the $y$-axis. The same results as for 1D systems are observed if the order parameter possesses zero longitudinal wavevectors ($k_y = 0$). In the case of $k_y \neq 0$ (see Fig.~\ref{fig3}a,b) the parity switch threshold ($P_\text{crit}$) with weak stochastic noise can result in the formation of localized defect states traveling along the channel carrying opposite parity charges (see Fig.~\ref{fig3}c,d,e). The phase pattern of the defects reveals that they're topologically distinct from the defect-free solution and thus they can be classified as topologically stable. We set our parameter values similar to previous works~\cite{keeling_spontaneous_2008}: $\alpha = 0.003$ ps$^{-1}$$\mu$m$^{2}$ , $R = 0.3\alpha$, $\Gamma = 0.1$ ps$^{-1}$, $m = 10^{-4} m_0$, $g_P = 3 \alpha$, where $m_0$ is the free electron rest mass. 
\begin{figure}[!t]
  \centering
		\includegraphics[width=0.5\textwidth]{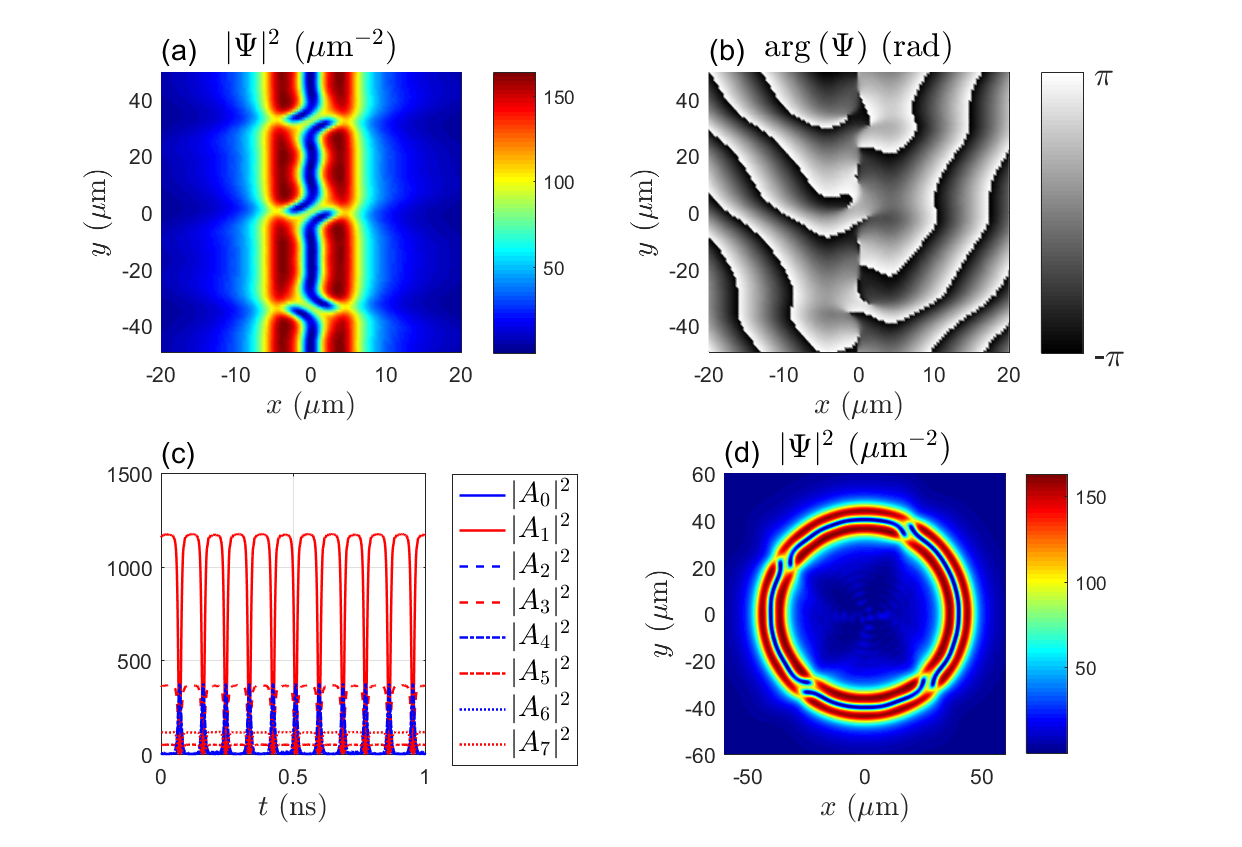}
	\caption{Condensate (a) density and (b) phase with three defect states after instantaneous switching of the pump at $P = 1$ ps$^{-1}$. (c) Mode population for $y=\mbox{const.}$ In the final state the defects are equidistant from each other and travel along the channel showing regular parity beats. (d) Defects traveling along a circular pump.}
\label{fig4}
\end{figure}

Instantaneous switching of the pump and subsequent condensation of the polaritons is a transition from a linear state to a saturated state which can also give birth to defects seeded by random noise. Nonresonant pulsing of the defect free channel can also induce their formation. In Fig.~\ref{fig4}a,b we show the final state after instantaneous switching of the pump where three defects travel in unison along the channel. Differently from Fig.~\ref{fig3} where the channel is of even parity with odd parity defects, here the channel is of odd parity with even parity defects. This result underlines that the defects can exist in either parity opposite of the channel. We observe defect formation at pump values where only one channel solution is stable verifying that the origin of the defects is not due to the bistability between the channel solutions~\cite{Helgi_SwitchWav_2015PRB}. We observe that the number of defects forming in the channel scales with the pump intensity at both $P_\text{cond}$ and $P_\text{crit}$ (see Sec.~\ref{sec.app3}). Formation of the defect domains scaling with the system control parameters (here $P$) is known as the Kibble-Zurek mechanism~\cite{Kibble_Topology_JPA1976,Zurek_Cosmo_Nature1985} and has been investigated for uniform pumping scenarios in polariton condensates~\cite{Matuszewski_Universality_PRB2014, Liew_NoneqDyn_PRB2015}.

\begin{figure}[!t]
  \centering
		\includegraphics[width=0.5\textwidth]{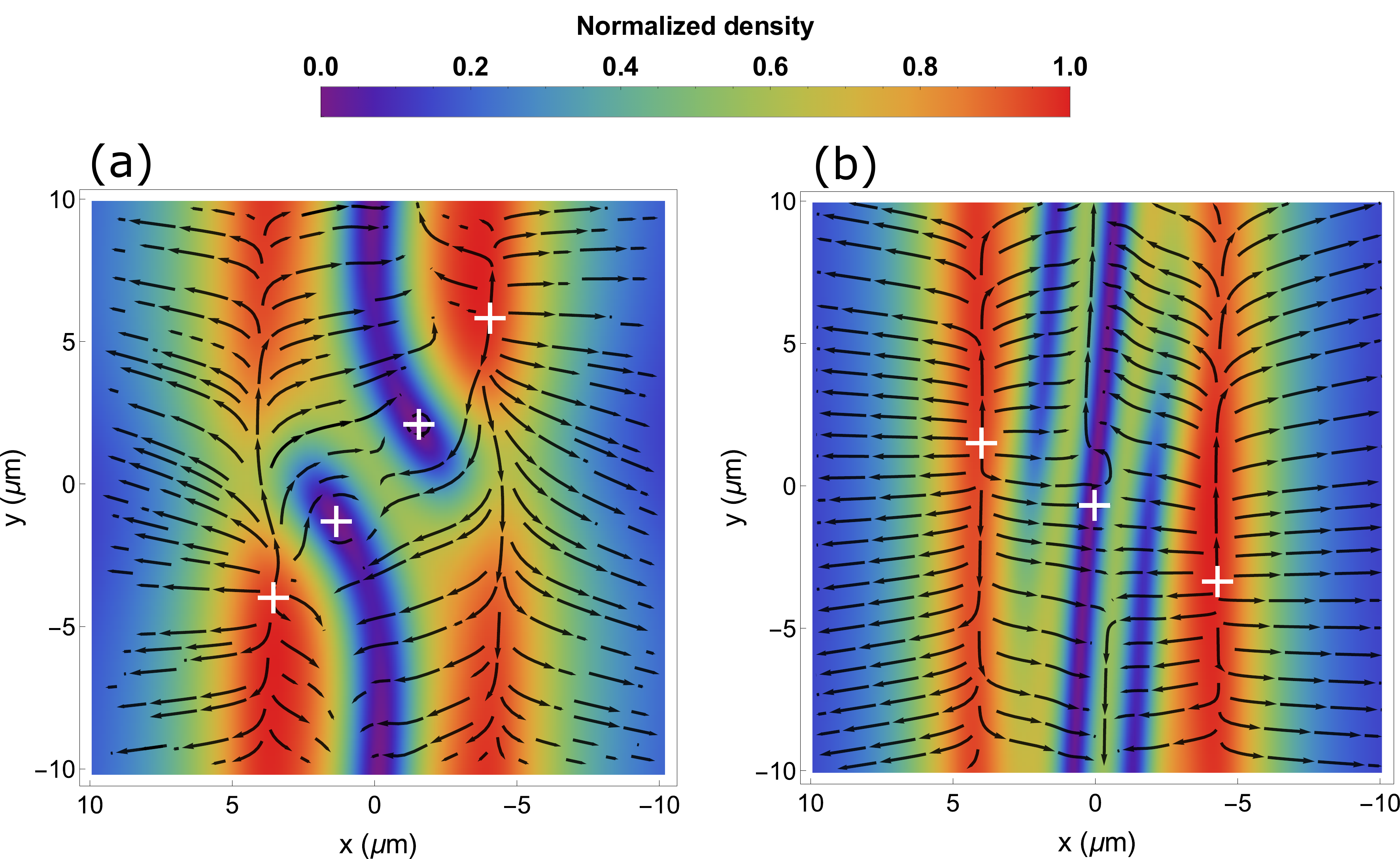}
	\caption{Velocity streamlines (black arrows) plotted over the condensate density (colormap) for an (a) even parity defect and (b) odd parity defect. White crosses denote sources of flux, circulation, and saddle points.}
\label{fig5}
\end{figure}

The defects are found to possess repulsive interactions and are unable to pass through each other. Instead, they display damped oscillatory behavior until maximum separation is achieved between the defects where they either become static or move in unison along the channel. If the channel ends are open the defects escape and dissipate. In order to capture the defects in experiment, a circular channel can be used (see Fig.~\ref{fig4}d). In Fig.~\ref{fig5} we plot the condensate velocity field (black arrows) and density (colormap) for the two different defect types. In Fig.~\ref{fig5}a we observe two in-phase sources of flux and two vortical points (white crosses) causing polaritons to flow freely across the channel. In Fig.~\ref{fig5}b we find two sources of flux $\pi$ out-of-phase and a single saddle point in the center of the channel. We note that the defect in Fig.~\ref{fig5}b can be regarded as a type of {\it dark soliton} due to the density minimum gashing diagonally across the channel. In Fig.~\ref{fig5}a the defects are reminiscent of {\it bright solitons} since a finite density connects across the channel. From the point of view of stability our solitons resemble more the infinitely propagating defects appearing under coherent pumping, which include domain walls in polariton neurons~\cite{Liew_Neurons_PRL2008}, bright solitons appearing near the inflection point of lower polariton branch~\cite{Egorov_BrigthSoliton_PRL2009} and parametric solitons~\cite{Egorov_PolSoliton_PRB2011}.

The coordinate dependent physics makes stability analysis infeasible but rescaling the order parameter $\Psi \to \Psi / \sqrt{R}$ shows that the nonlinear physics of the system depend only on $\alpha/R$. Starting from an odd parity channel (Fig.~\ref{fig3}a) populated by a single traveling defect we numerically resolve the boundary of stable parity solitons in the $P$-$\alpha/R$ parameter plane (see Sec.~\ref{sec.app4}). Expectedly, the regime of stable defects depends also on their velocity making current observations non-exhaustive. For small $\alpha/R$ and $P$ the soliton is unable to hold together the supercurrents and breaks up. At high $\alpha/R$ and $P$ a formation of vortex-antivortex pairs appears along the guide, modulating the density but not destroying the original soliton. These vortex-antivortex pairs have been reported before in the stability analysis of dark soliton stripes (akin to Fig.~\ref{fig3}a)~\cite{Smirnov_DarkSolitons_PRB2014} and observed experimentally~\cite{Grosso_VortexStreets_PRL2011}. For increasing $P$ one eventually hits $P_\text{crit}$ where both guide and existing solitons undergo a dramatic change. Here the domain walls of the soliton spread apart to fill up the guide, effectively switching the parity of the entire guide (see animation 2 in the Supplemental Material~\cite{supplemental}). For a finite guide the soliton would then spread out and vanish (domain walls exit the system). However, for a closed (periodic) guide the domain walls meet again forming a now a soliton of the opposite parity preserving the soliton above $P_\text{crit}$.

Lastly, we investigated the propagation of the defects in the presence of static disorder~\cite{Savona_disorder_PRB2006}. Depending on the disorder landscape the defects can either pass unhindered along the channel, become trapped between disorder maxima, or break up. To our surprise, we find that a defect trapped in the potential landscape oscillates without damping as opposed to damped collisions with other defects (see animation 1 in the Supplemental Material~\cite{supplemental}).

\section{Conclusions}
We've analyzed analytically and numerically the effects of symmetric nonresonant pumping in polariton condensates in 1D systems. We show that the minimum condensation threshold belongs to a condensate of definite parity. We also show the existence of a second critical pump power where the phase degree of freedom allows the opposite parity solution to take up the gain and drive the other parity to zero. Stretching such symmetric pump profiles to form channels in 2D system allows the formation of solitonic parity-defects. The defects exist over wide range of parameter values including no external or pump-induced trapping. The defects possess nontrivial velocity patterns making them topologically distinct, and display damped collisions when approaching one another. We stress that our defect states travel persistently along the pump channel and do not decay even with  constant dissipation present in the system, unlike what would happen if one tried, for example, to excite a traditional dark soliton in a 1D channel geometry.

\section{Acknowledgements}
H.S. and I.S. acknowledge support by the Research Fund of the University of Iceland, The Icelandic Research Fund, Grant No. 163082-051. T.L. was supported by the Singaporean MOE grants No. 2015-T2-1-055 and No. 2016-T1-1-084. I.A.S. acknowledges support from a mega-grant No. 14.Y26.31.0015 and GOSZADANIE No. 3.2614.2017/4.6 of the Ministry of Education and Science of Russian Federation.

\appendix
\section{EXTENSION TO RESERVOIR MODELS} \label{sec.app1}
\begin{figure}[!b]
  \centering
		\includegraphics[width=\linewidth]{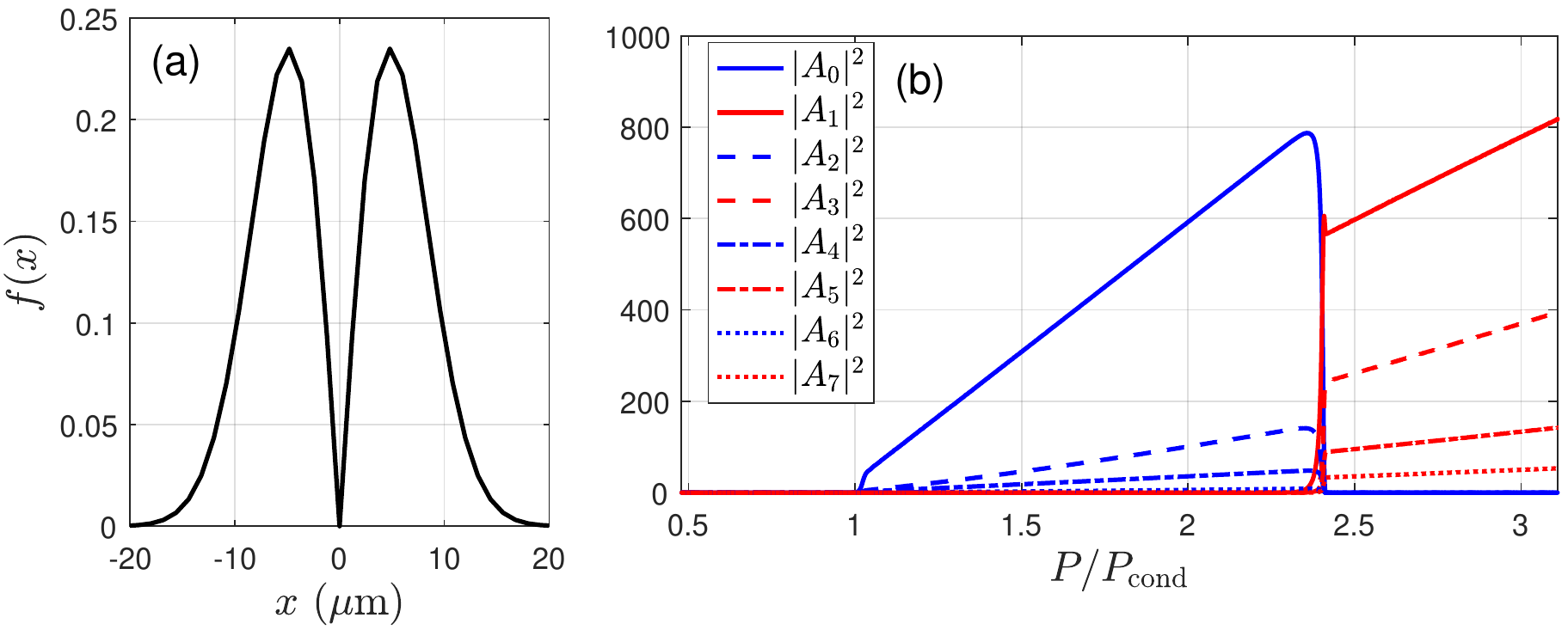}
	\caption{(a) Example choice of pump profile producing the parity switch by propagating Eq.~\ref{eq.GPR}. (b) Population of the condensate in the first eight linear basis eigenstates.}
\label{fig6}
\end{figure}

The transition from one parity to another is not exclusive to Eq.~(\ref{eq.GP0}) in the main text which describes static gain at the pump location at all times. An alternative model describes the gain of polariton through a dynamical reservoir $n_R$ of hot excitons governed by a rate equation~\cite{wouters_excitations_2007},
\begin{align} \label{eq.GPR}
i \dot{\Psi} & = \left[ - \frac{\hbar}{2m} \frac{\partial^2}{\partial x^2} + \frac{i}{2} \left(  R n_R - \Gamma \right) \right] \Psi, \\
\dot{n}_R & = - (\Gamma_R + R |\Psi|^2) n_R + P(x).
\end{align}
Here $\Gamma_R$ is the reservoir dissipation rate and $R$ now plays the role of in-scattering rate from the reservoir to the condensate. We note that under the current consideration the effective potential from interactions with polaritons and the reservoir are set to zero. The reservoir steady state is given by,
\begin{equation}
n_R(x) = \frac{P(x)}{\Gamma_R + R |\Psi(x)|^2},
\end{equation}
where $P(x) = P f(x)$. This more complicated expression of the polariton gain mixes the parities of the system and the pump elements $f_{nm}$ no longer vanish for $n$ and $m$ of different parities. A good choice of a pump profile $f(x)$ can however reproduce the switch in parities (see Fig.~\ref{fig6}) underlining that the transition is still present in reservoir models.

\section{EFFECTS OF NOISE AND ASYMMETRY IN DRIVING FIELD} \label{sec.app2}
All results presented are done using a stochastic low amplitude initial condition and background noise added to the order parameter at timesteps much smaller then the characteristic polariton timescales. The noise serves as a method of breaking system symmetries and giving rise to nontrivial states at critical transition points such as condensation ($P_\text{cond}$) and parity switching $(P_\text{crit})$. 

\begin{figure}[!t]
  \centering
		\includegraphics[width=\linewidth]{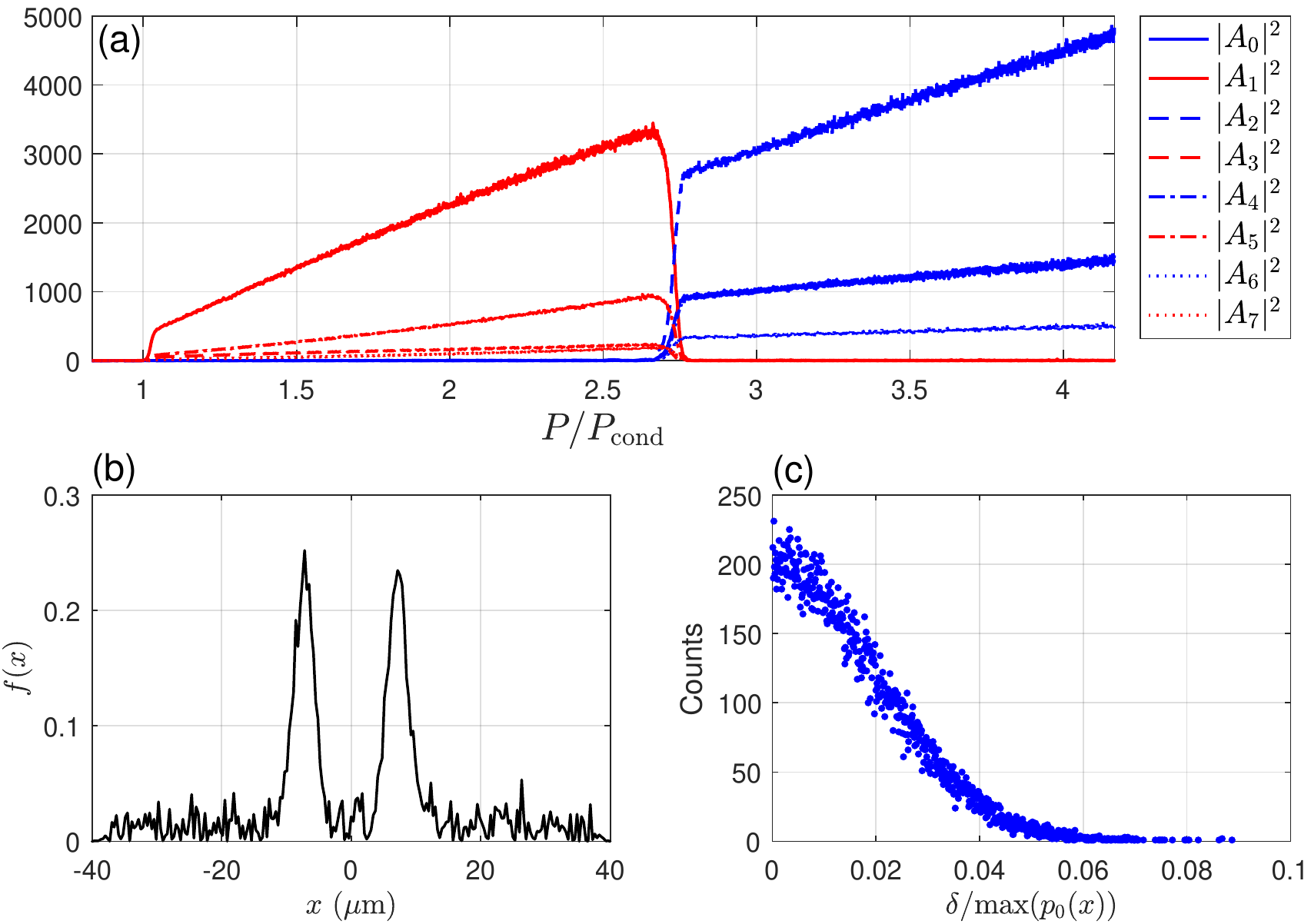}
	\caption{(a) Population of the first eight linear basis eigenstates showing a change in parity analogous to Fig.~2a in the main text. (b) Snapshot of the noisy pump profile. (c) Distribution of the noise.}
\label{fig7}
\end{figure}

We now investigate additional Gaussian distributed noise field with zero mean added to the pump profile $f(x)$ at small time steps $\tau \ll \Gamma^{-1}$. 
\begin{equation}
f(x) = f_0(x) + \delta(x,\tau).
\end{equation}
Here $f_0(x)$ is the unperturbed pump shape. This tests the sensitivity of the parity dependent nature of the gain-decay mechanism. Fig.~\ref{fig7}a shows the parity switching taking place undeterred by the static noise distribution (Fig.~\ref{fig7}c) added to the pump profile (Fig.~\ref{fig7}b).

\begin{figure}[!t]
  \centering
		\includegraphics[width=\linewidth]{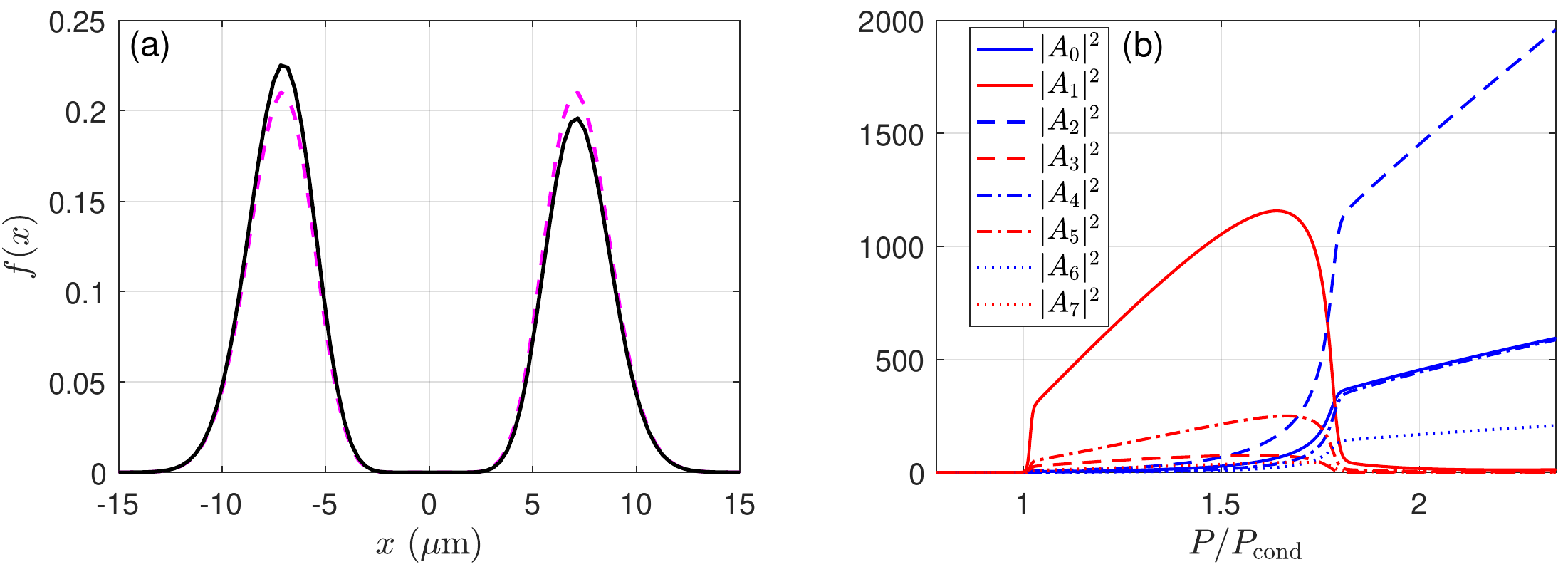}
	\caption{(a) Skewed (black solid line) vs symmetric (dashed magenta line) pump profile. (b) Population of the first eight linear basis eigenstates showing a gradual change in parity for the asymmetric pump profile in panel (a).}
\label{fig8}
\end{figure}

We next check static asymmetry in the pump profile. Results in Fig.~\ref{fig8} show that asymmetric profiles alters the evolution of the condensate as a function of pump power $P_0$. The transition takes place over a larger interval of pump powers due to the pump now mixing different parity eigenstates. Regardless, the for reasonable skewing of the pump profile (Fig.~\ref{fig8}a) the parity transition still takes place where a new solution becomes dominant and quenches the other (Fig.~\ref{fig8}b). The results underline that the physics at play are robust against reasonable amounts of noise and skewing which can be expected in experiments. We note that here we have only investigated one type of a pump profile whereas other profiles with a different set of pump elements $f_{nm}$ might be less affected by asymmetry.

\section{KIBBLE-ZUREK TYPE SCALING} \label{sec.app3}
We investigate the number of defects/solitons appearing as a function of pump intensity increased rapidly above condensation threshold ($P_\text{cond}$) and parity switching threshold $(P_\text{crit})$. It has been shown that the standard Kibble-Zurek theory, which captures the scaling of the defect number as a function of control variables (e.g. temperature), does not apply to nonequilibrium systems such as exciton-polariton condensates described by the complex Ginzburg-Landau equation~\cite{Matuszewski_Universality_PRB2014}. However, the system is still described by competitive timescales of gain-and-decay and can display scaling between pump intensity as a control parameter and defect formation~\cite{Liew_NoneqDyn_PRB2015, Solnyshkov_KZmicropillar_PRL2016}. 

\begin{figure}[!t]
  \centering
		\includegraphics[width=1.0\linewidth]{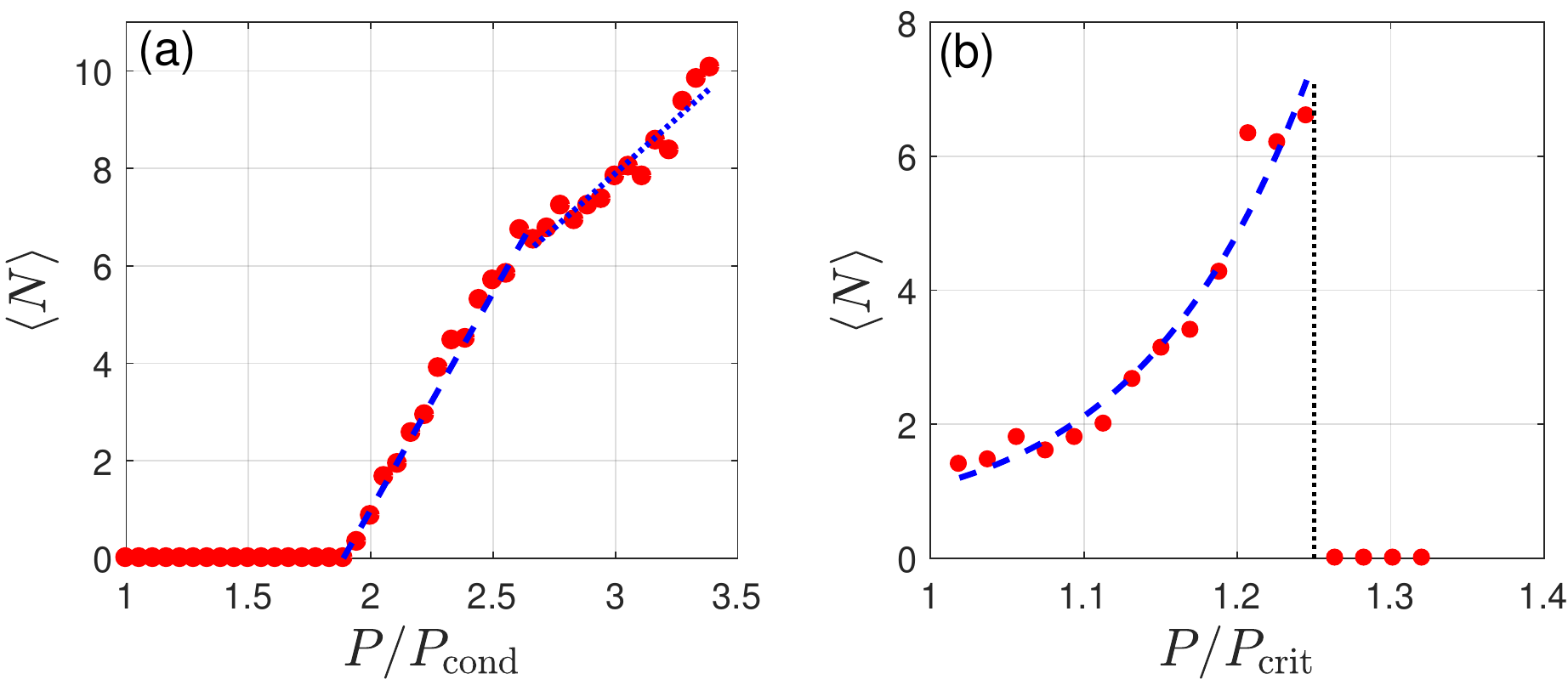}
	\caption{(a) Average number of defects $\langle N \rangle$ versus instantaneous switching of the pump intensity from an uncondensed state. (b) $\langle N \rangle$ versus instantaneous increase in pump intensity around $P_\text{crit}$ triggering a change from an odd parity condensed state to an even parity state. Blue dashed lines are guides for the eye. In both panels $R = 0.3\alpha$ and $g_P = 0$.}
\label{fig9}
\end{figure}

\begin{figure}[!t]
  \centering
		\includegraphics[width=\linewidth]{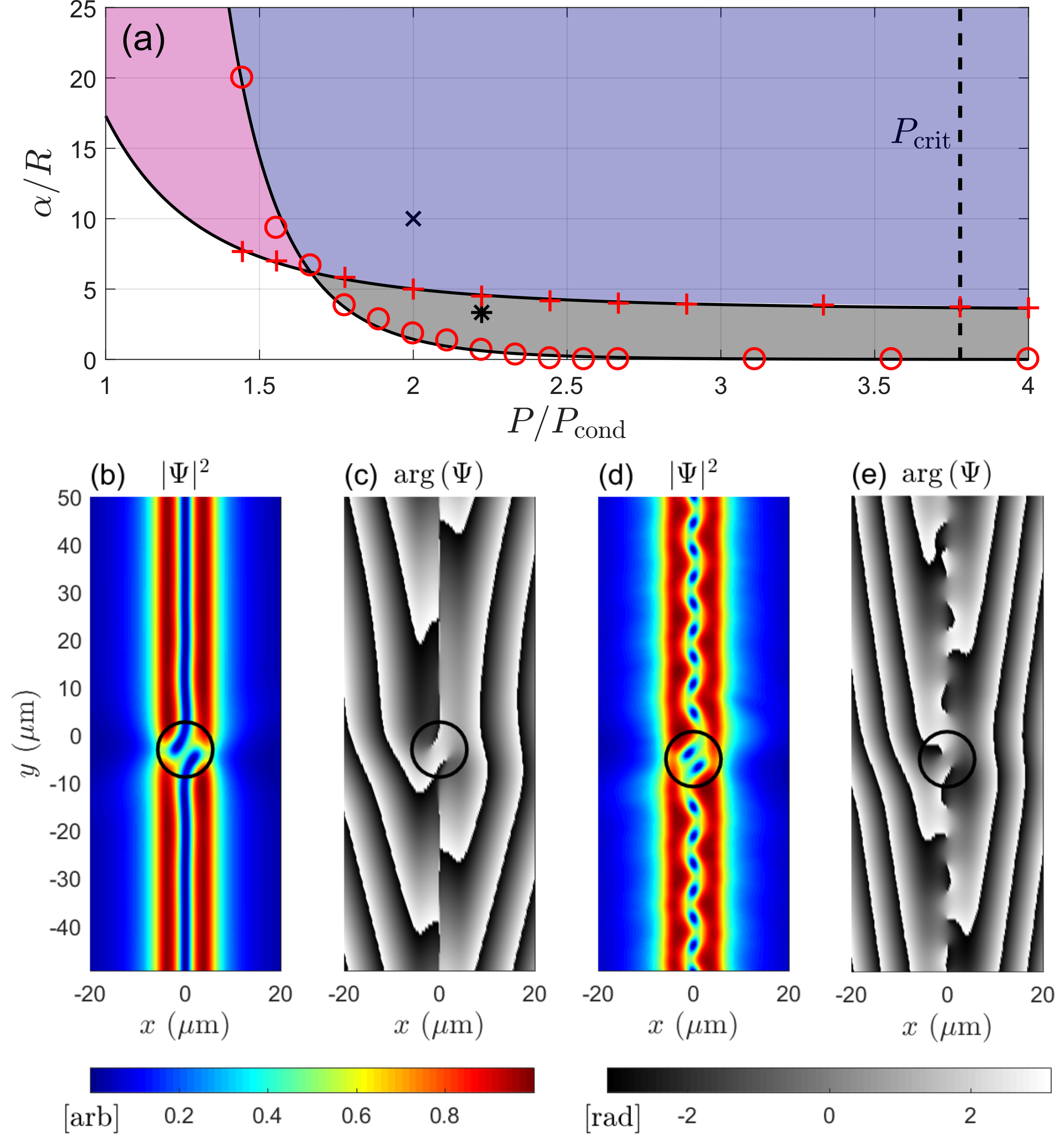}
	\caption{(a) Numerically estimated boundaries of stable defects and condensate instability. Above red circles stable solitons exist in the channel. Above red crosses vortex-antivortex formation takes place along the channel. The black solid lines are a power law fit of the data. Different regimes have been colored for clarity. Black x-mark correspondd to panels (b,c) and asterix to (d,e). (b-e) Condensate density and phase maps where the soliton location in the channel is marked by a black circle. Here $g_P = 0$.}
\label{fig10}
\end{figure}

We propagate Eq.~(\ref{eq.GP0}) with weak white noise order parameter ($|\Psi| \sim 0$) initial condition and with a set $P$ value (i.e., instantaneous activation of the driving field intensity). The average defect number $\langle N \rangle$ is defined here as the number of observed parity solitons in a 100 $\mu$m long condensate guide (periodic boundaries), averaged over 30 trials. We set $\alpha/R = 3$ which, according to Fig.~\ref{fig10}, permits stable defects above $P \approx 1.8 P_\text{cond}$. Indeed, around this value a linear scaling of the defect number takes place (see Fig.~\ref{fig9}a). Around $P \approx 2.6 P_\text{cond}$ the growth of the defect number changes which is associated with the guide having reached its maximum number of unmodulated defects. The soliton size is approximately 13 $\mu$m (see Fig.~\ref{fig4}a) over a wide range of pump values. Since $8 \times 13 \ \mu\text{m} > 100 \ \mu\text{m}$ we infer that $\langle N \rangle = 7$ is the smallest number of solitons to be contained in the guide without becoming distorted by the presence of other solitons and affects the rate of defects appearing at higher pump powers.

More interestingly, instantaneous increase of the pump intensity above $P_\text{crit}$ where the parities of the condensate switch places also results in defect formation. Using a defect free condensate at $P = P_\text{in} < P_\text{crit}$ as an initial condition, we shock the system by suddenly changing to $P > P_\text{crit}$ allowing the formation of defects around the switch (see Fig.~\ref{fig9}b). Unlike condensing the system suddenly, the defect formation here is due to the singular phase behavior where the order parameter goes to zero (e.g., at $x=0$ for odd parity state). The defects are then seeded from stochastic fluctuations when the parity of the condensate is changing. Given our initial condition, if the pump is increased too much the condensate overshoots the parity switch interval and no defect formation is observed (indicated by black dotted line in Fig.~\ref{fig9}b). 

We stress that weak white noise was introduced at small timesteps ($t \ll \Gamma^{-1}$) to mimic classical thermal fluctuations. The heuristic employed here serves to demonstrate, as expected, an increase in defect formation with greater jumps in pump intensity although methods relying on the truncated Wigner approximation and a stochastic set of equations~\cite{Wouters_PRB2009, Read_PRB2009} would more accurately bring out the scaling laws at play.

\section{DEPENDENCE ON PUMP POWERS AND NONLINEARITIES} \label{sec.app4}
By scaling the order parameter it can be shown that the nonlinear physics of the system depends only on the ratio $\alpha/R$. For a given pump profile $f(\mathbf{r})$ (same as in Figs.~\ref{fig3}-\ref{fig5}) we investigate the regime of stable parity solitons as a function of pump power $P$ and nonlinearity (see Fig.~\ref{fig10}a). Here, we set $g_P = 0$ since it doesn't play an important role in the stability of the defects. Above the red circles the solitons are stable within the condensate. Note that $\alpha \neq 0$ is necessary in order for the solitons to stay stable through interactions. The flattening on the horizontal axis corresponds to the increased density of the condensate requiring smaller values of $\alpha$ to produce the same net nonlinear effect to stabilize the solitons.

Above the red crosses the condensate starts becoming turbulent through small fluctuation and forms a train of vortices and antivortices interchangeably such that the net rotation in the condensate is still zero (see Fig.~\ref{fig10}d,e). The boundaries in Fig.~\ref{fig10}a are resolved by using a single even parity soliton populating the guide as an initial condition and then adiabatically tuning the parameters until the state changes. Black x-mark refers to Figs.~\ref{fig10}b,c and asterix to to Figs.~\ref{fig10}d,e where the solitons are marked by black circles. Remarkably, moving the soliton into the turbulent regime (from the asterix to the x-mark) does not result in its destruction but rather it remains as a chink in the chain of vortices and anti-vortices.

\bibliography{Bibliography}

\end{document}